\documentstyle[12pt,aasms4]{article}

\newcommand{\beq}{\begin{equation}}
\newcommand{\eeq}{\end{equation}}
\newcommand{\Mdot}{\dot{M}~}

\newcommand{\kms}{\mbox{ km s$^{-1}$}~}

\newcommand{\Moy}{\mbox{M$_{\odot}$ yr$^{-1}$}~}

\newcommand{\Ro}{\mbox{R$_{\odot}$}}

\begin{document}

\def\etal{{\it et~al.\ }}
\def\eg{{\it e.~g.\ }}
\def\ie{{\it i.~e.,\ }}

\title{LATE AGB MAGNETIC CYCLES: MHD SOLUTIONS FOR THE HST PN RINGS} 

\author{Guillermo Garc\'{\i}a-Segura\altaffilmark{1}, Jos\'e Alberto
 L\'opez\altaffilmark{2}, Jos\'e Franco\altaffilmark{3,4}}
\affil{Instituto de Astronom\'{\i}a-UNAM, Apdo Postal 877,
Ensenada, 22830 Baja California, Mexico}

\altaffiltext{1}{Email address: ggs@astrosen.unam.mx}
\altaffiltext{2}{Email address: jal@astrosen.unam.mx}
\altaffiltext{3}{Instituto de Astronom\'{\i}a-UNAM, 
Apdo Postal 70-264, 04510 M\'exico D. F., Mexico}
\altaffiltext{4}{Email address: pepe@astroscu.unam.mx}

\begin{abstract}
\noindent

The Hubble Space Telescope has revealed the existence of multiple, 
regularly spaced, and faint concentric shells around some planetary 
nebulae. 
Here we present 2(1/2)D magnetohydrodinamic numerical simulations of the
effects of a solar-like magnetic cycle, with periodic polarity inversions,
in the slow wind of an AGB star. The stellar wind is modeled with a 
steady mass-loss at constant velocity. This simple version of a 
solar-like cycle, without mass-loss variations, is able to reproduce many
properties of the observed concentric rings. The shells are formed by 
pressure oscillations, which drive compressions in the magnetized wind. 
These pressure oscillations are due to periodic variations in the field 
intensity. The periodicity of the shells, then, is simply a half of the 
magnetic cycle since each shell is formed when the magnetic pressure
goes to zero during the polarity inversion. As a consequence of the 
steady mass-loss rate, the density of the shells scales as $r^{-2}$, and 
their surface brightness has a steeper drop-off, as observed in 
the shells of NGC 6543, the best documented case of these HST rings. 
Deviations from sphericity can be generated by changing 
the strength of the magnetic field. For sufficiently strong fields, a 
series of symmetric and equisdistant blobs are formed at the polar axis, 
resembling the ones observed in He 2-90. These blobs are originated by 
magnetic collimation within the expanding AGB wind.

\end{abstract}


\vfill\eject
\section{Introduction} 

The multiple concentric rings, or arcs, that were recently discovered by 
the Hubble Space Telescope around a handful of planetary nebulae (PNe)
is one of the most puzzling and unexpected results delivered by the HST
(see Kwok, Su, \& Hrivnak 1998; Hrivnak, Kwok \& Su 2001; Terzian \& Hajian
2000 and references 
therein). The best documented case of these systems of faint concentric 
rings (hereafter called HST rings) is displayed by NGC 6543 (Balick, 
Wilson \& Hajian 2000). In reality, they are regularly spaced concentric
shells, indicating quasi-periodic events with time intervals, assuming 
typical expansion velocities of AGB winds, in the range of 500 to 1500 
years. These time scales are too short for thermal pulses and too long 
for acoustic envelope pulsations and, thus, the origin of the rings 
cannot be ascribed to any of this type of events. Soker (2000) made a 
critical review of the mechanisms that have been proposed to explain 
them, and he indicates that mass-loss variations associated with 
solar-like magnetic cycles are perhaps the best alternative for their 
origin. 

In a more recent paper, Simis, Icke \& Dominik (2001) discuss a new
non-magnetic alternative, and present detailed 1D hydrodynamical
simulations for the acceleration of a dusty AGB wind. They include dust 
grain nucleation and growth in the AGB atmosphere, and the dust and gas
components are treated as two separated fluids, without assumptions
about grain drift or dust-gas coupling. Recently formed grains are
accelerated first and then, as they drift throught the gas at relatively
large speeds (sometimes well above the equilibrium drift velocity), the 
drag force transmits the momentum to the gas. For non-magnetic dusty 
flows, and depending on grain properties (\ie chemical composition, 
size, and charge) and local conditions (\ie ionization fraction and 
temperature), an efficient momentum transfer requires that the grains 
sweep throught gas column densities of the order of $N \sim 10^{19} a_5$
cm$^{-2}$ (where $a_5$ is the grain radius in units of $10^{-5}$ cm; 
Franco \etal 1991). Thus, the grain drift velocity can be larger than 
its equilibrium value for a certain distance, compressing the gas at
some locations and creating regions with larger dust-to-gas mass 
ratios. These compressions appear in a cyclic manner in their 1D
simulations and, if such a flow configuration is stable in two or three 
dimensions, they can drive a variable mass loss rate and may lead to
the creation of dusty shells. The stability of this process, however, is 
difficult to explore at the moment, and it is unclear if the compressions 
can truly create shells or if they only lead to inhomogeneities in these
radiatively driven outflows. Thus, at present, one can consider that dusty
flow oscillations (if they really exist) or solar-like
magnetic cycles are two possible candidates to generate the HST rings.

In Soker's (2000) view, the magnetic field plays no direct role in the 
evolution of the AGB wind, however temporal variation in the number of 
magnetic spots would be able to modify the mass-loss rate. The number of cool
spots over the AGB surface, which could be preferred sites for dust formation,
is controlled by the magnetic cycle. Thus, given that the mass-loss is 
driven by radiation pressure on dust grains, the same cycle may also 
regulate periodic variations in the mass-loss rate. In this 
interpretation, the magnetic field is a passive player with no dynamical
effects, and the mass-loss rate simply follows the spot cycle activity. 
In addition, to make a logical association with bipolar PNe, he suggests
that a stronger magnetic activity could be expected from dynamo 
amplification in binary systems. 

Here we develop a different point of view and explore some of the actual
dynamical effects of a solar-like magnetic cycle. The possibility of a
solar-like magnetic dynamo at the AGB phase has been recently discussed by
Blackman \etal (2001), and they conclude that dynamo amplification is
likely to operate in rotating AGB stars. A logical extension of this result
is that a solar-like activity, including dynamo and cycles, is also expected
in some AGB stars. We thus build a very simple model, without mass-loss 
variations, and perform 2(1/2)D magnetohydrodynamic computations 
considering a cyclic polarity inversion of the surface magnetic field of an
AGB star. Our results show the importance of MHD effects in the formation 
of the HST rings and successfully reproduce their main features. This 
indicates that modulated mass-loss episodes are not really necessary to 
generate the rings. In Section 2 we describe the MHD models and results. A 
brief discussion is given in Section 3.

\section{Numerical Models} 

\subsection{The Method}

The simulations have been performed using the magnetohydrodynamic code 
ZEUS-3D (version 3.4), developed by M. L. Norman and the Laboratory for 
Computational Astrophysics. This is a finite-difference, fully explicit, 
Eulerian code descended from the code described in Stone \& Norman 
(1992). A method of characteristics is used to compute magnetic fields, 
as described in Clarke (1996), and flux freezing is assumed in all the 
runs. We have used spherical polar coordinates ($r,\theta,\Phi$), with 
reflecting boundary conditions at the equator and the polar axis. 
Rotational symmetry is assumed with respect to the rotational (polar) 
axis, and our models are effectively two-dimensional. The simulations are
carried out in the meridional ($r$, $\theta$) plane, but three 
independent components of the velocity and magnetic field are computed 
(\ie the simulations are ``two and a half'' dimensions). We are unable 
to include, as Simis \etal (2001) have done with their two-fluid code,
the effects of dust in our simulations.

Our grids consist of $200 \times 180$ equidistant zones in $r$ and 
$\theta$, respectively (with a radial extent of 0.1 pc, and an angular 
extent of $90^{\circ}$), and the innermost radial zone lies at $r=2.5 
\times 10^{-3}\,$pc from the central star. These values are used in 
all the simulations shown in Figures 1 -- 3.

The equations for the stellar wind flow are simple, since we model an
isothermal, slow AGB wind with constant mass-loss rate ($10^{-6}$ \Moy) 
and radial velocity (10 \kms). These values represent the boundary 
conditions used in the first five innermost radial zones. Rotational 
effects are not important in these outflows, since we have used a small 
value for the stellar rotation velocity, 0.01 \kms (see 
Garc\'{\i}a-Segura \etal 1999 for more details).

The novel aspect in this paper is the simple treatment of the stellar
magnetic field ($B_{\rm s}$), which it is allowed to change sign in a
cycle of the form:
\beq
B_{\rm s}(t) = B_{\rm max} \cos (2 \pi \frac{t}{P}),
\eeq
where $B_{\rm max}$ is the maximum average $B$-field at the AGB surface,
and $P$ is the period of the magnetic cycle. Since we do not know the 
true variation form of the field, this functional form is just a first 
simple approximation. As in the case of the Sun, we assume that $B_{\rm 
max}$ has opposite signs at each hemisphere, with a neutral current
sheet near the equatorial plane (\eg Wilcox \& Ness 1965; Smith,
Tsurutani \& Rosenberg 1978). Its average thickness in the solar case
is of about $10^8$ cm, and its presence does not affect the field
outside the equatorial sections. For simplicity, given that we compute
only one hemisphere, we neglet the size of this current sheet.
The equation for the inner boundary wind toroidal field on the 
computational domain, which is restricted to 
one hemisphere ($ 0 \leq \theta < \pi/2 $), is:
\beq
B_{\phi}(t) = B_{\rm s}(t) \frac{v_{\rm rot}}{v_{\infty}} 
\left( \frac{ R_{\rm s}}{r} \right)^2 \left( \frac{r}{ R_{\rm s}} - 1 
\right)
{\rm sin}\ \theta ,
\eeq
where $v_{\rm rot}$ is the stellar rotation velocity, $v_{\infty}$ the 
wind velocity, and $R_{\rm s}$ the stellar radius. The function $\sin 
\theta$ cancels the toroidal component at the symmetry axis, $\theta =0$
(pole). The poloidal field component can be neglected at large 
distances, so that our field configuration naturally satisfies the 
condition $\nabla \cdot  B = 0$.
 
The parameter which identifies our models is the ratio of the magnetic 
field energy density to the kinetic energy density in the wind (Begelman 
\& Li 1992)
\beq 
\sigma (t)= \frac{B^2}{4 \pi \rho v_{\infty}^2} = \frac{B_{\rm s}^2(t) 
R_{\rm s}^2}{
\Mdot v_{\infty} } \left( \frac{ v_{\rm rot}}{v_{\infty}} \right)^2 \,\,.
\eeq
Obviously, this parameter is always positive and oscillates with twice 
the frequency of the stellar magnetic cycle.

\subsection{Results}

The first model, shown in Figure 1, corresponds to the case of a slow and
dense wind ($v_{\infty}=10 \kms$ and $\Mdot = 10^{-6} \Moy$), in which 
the magnetic field at the surface of the AGB star cycles with a full 
period of 2000 yr. A peak value $\sigma_{\rm max}=0.01$ is used for a 
slowly rotating AGB star ($v_{\rm rot} = 0.01$ \kms), which is achieved,
for example, when the star has $B_{\rm s}=53$ G and $R_{\rm s}=1$ AU,
or $B_{\rm s} = 113 $ G and $R_{\rm s} = 100 $ \Ro. For the cases 
discussed by Soker (2000), $R_{\rm s} = 2 $ AU and $ 0.2 \kms \leq
v_{\rm rot} \leq 2 \kms $, the required surface magnetic fields are 
within 1.3 G $\geq B_{\rm s} \geq $ 0.13 G. These field strenght values
are modest compared withn the ones that could be achieved by dynamo
amplification during AGB evolution (Blackman \etal 2001).

Figure 1 displays radial cuts, on the equatorial plane, of some of the
wind variables at $t= 10,000$ yr after the onset of the magnetized wind. 
The first (top) panel shows the existence 
of equidistant concentric shells, separated by 1,000 yr, a half of the
magnetic cycle period. Each density peak corresponds to a minimum in the
magnetic pressure (second panel), which also occurs at the moment when 
the magnetic field changes polarity (fourth panel). The last panel of
Figure 1 displays the radial velocity, which clearly shows the gas 
response to the local variations in the total pressure (third panel)
 of the wind. 

Thus, shell formation in a magnetized wind with variable field strength 
is a straightforward process. The outflowing plasma notices the magnetic 
pressure depressions, both upstream and downstream in the wind, and moves
towards the low-pressure sites. The wind is then compressed at these 
locations, increasing the local density, to compensate for the low
magnetic pressure values. Obviously, the particular values achieved at 
these density peaks, and the density constrast between ring and 
inter-ring zones, depend on the model assumptions. Thus, these values can
be modified by changing the amplitude of the pressure fluctuations (\ie
by changing the maximum field intensities, or adding mass-loss or 
velocity variations, etc.) but, for our purpouses, it is sufficient to 
show that a reasonable density constrast is achieved by this simple
model. The total pressure plot (third panel), shows that the local 
pressure fluctuations decrease with time (or position in the wind). As 
the plasma flows from the positions of the magnetic peaks and compresses
the gas of the valleys, a series of MHD waves are continuosly driven that
maintains small gas oscillations in the expanding wind.

Figure 2 shows the emission measure of four different models, with 
$\sigma_{\rm max} = 0.001, 0.01, 0.05, 0.1 $. The plots nicely reproduce 
the typical spacing of the rings, and show the general drop-off 
expected for a constant mass-loss wind. The structures are in reality a 
set of spherical or quasi-spherical hemispheric shells, and their
projected maximum column densities in the plane of the sky give the 
impression of concentric rings. For large enough values of $\sigma_{\rm 
max}$, the magnetized AGB wind is able to self-collimate towards the 
polar axis. This self-collimation is also present in the free streaming 
fast winds that form some PNe (see Garc\'{\i}a-Segura \etal 1999) but, 
given the large densities of the slow AGB winds, the self-collimated 
structures are particularly prominent, and easily observable features 
in this case. The two right panels in Figure 2 clearly show the 
resulting structures, that are better defined and more conspicous
in the $\sigma_{\rm max} = 0.1$ case. The self-collimated gaseous 
structures also follow the periodic variations of the magnetic field, 
resulting in two strings of regularly spaced blobs located along the 
polar axis resembling those observed in He 2-90 (Sahai \& Nyman 2000).

\section{Discussion and Summary}

A relevant point for the applicability of this model is whether or not 
the $B$-field can be considered frozen into the outflowing dusty AGB 
wind, and if the simple MHD effects that we have described are truly 
operative during wind evolution. This question has been previously 
explored by several authors and in much more restrictive environments. 
For instance, in collapsing molecular clouds with ionization fractions 
as low as $10^{-7}$, where ion drift could lead to substantial magnetic 
flux leakage. Nontheless, even at these very low ionization fraction 
values, numerical models of cloud collapse show that the field remains 
nearly frozen into the cloud for long periods of time (\eg Black \& 
Scott 1982). For AGB winds the situation is certainly more favorable 
because the ions are  provided by a number of species with low 
ionization potentials (Habing 1996; Cox 1997), and charged dust grains
are also well coupled to the magnetic field of the outflow. Thus, flux 
leakage is not considered important during the AGB phase, which occurs
during time scales of the order of 10$^3$ to 10$^5$ yr (for cases where 
ambipolar diffusion is important see Mac Low \etal 1995 and references 
therein). Later on, the AGB wind is photoionized by the central star of 
the PN (as in NGC 6543 or Hb 5), and the temperature and ionization 
fraction of the outflowing plasma is suddenly increased. Some pressure 
readjustments occur within the photoionized plasma at this time, but the
basic shell structuring remains unchanged. Obviously, flux leakage 
becomes irrelevant after this moment, and the rings can survive for 
longer periods of time.

An important issue related with the validity of the present results is 
whether or not magnetic reconnection can occur in the flow, modifiying 
some of the features appearing in the models. As stated before, in 
analogy with the solar case, it is likely that the general AGB $B$-field
has a dipolar structure, and a neutral current sheet should be present 
around the equatorial plane of their magnetized winds. The process
creates a pair of half-shells at each cycle, one per hemisphere and
separated by the neutral current sheet at the equatorial plane. Each
half-shell, in turn, is compressed by magnetic intershell regions with
opposite polarities. This implies that the compressed half shells also
act as neutral current sheets, separating regions with oppositely directed
field lines. Our simulations assume well ordered fields and the results do
not show, even in the more magnetic models, any signs of reconnection
within the shells. If one assumes that a twisted random field component
could be also present, a faster reconnection mode could also occur (\eg
Lazarian \& Vishniac 1999). Given the low field values in the shells,
however, any random field reconnection that may occur within them is
severily limited by the corresponding low Alv\'en speeds, and the energy
and flux involved in such a process has to be neglegible.

Another related question is if the shells can be formed only by 
modulated mass-loss rate episodes, without magnetic pressure in the 
wind, as proposed by Soker (2000) and Simis \etal (2001) (keeping in 
mind that the Simis \etal results could be unstable in 2D or 3D,
precluding the actual formation of a shell). In these cases, the
periodic formation of shells in the expanding wind is only due to the 
increase in mass-loss, and the higher shell densities can be maintained 
as long as the wind temperature decreases in these same locations. This 
is a perfectly reasonable possibility during the AGB phase. However, 
once the wind becomes photoionized, the plasma temperature becomes 
fairly homogeneous, and the rings tend to be washed away in a sound 
crossing time. Thus, ring survival is severely compromised in these 
purely hydrodynamic cases.

There is one additional observational element that suggests that the
magnetohydrodynamic mechanism is indeed relevant in these cases. Recent HST
images of He 2-90 show a series of brigth knots which have been interpreted
as due to symmetric and well collimated fast jets (Sahai \& Nyman 2000). New
spectroscopic observations by Guerrero \etal  (2001) have found that these
jet-like features are moving away from He 2-90 at a remarkably constant
radial velocity of only 26 \kms . Obviously, this radial speed value has to
be corrected for the inclination angle with respect to the plane of the sky.
This angle is unknown, but assuming that the axis of the jet-like features 
is located almost on the plane of the sky, they find an upper bound to the 
actual space velocity of about 290 \kms . However, the lack of any
trace of collisional excitacion within the flow or interaction with the 
ambient medium (\ie the sulfur and oxygen lines in the collimated 
structures are weak, and there are no signs of the bow-shocks expected for 
the leading parts of supersonic jets), makes it difficult to ascribe these 
highly collimated structures with a supersonic jet (this is reinforced by 
the observed narrow line profiles and lack of acceleration within the 
structures). In contrast, the constant and moderate velocitie values, and the 
similarity of the observed knotty structure with the largest $\sigma_{\rm 
max}$ model presented here provides a very atractive alternative for its 
origin. Figure 3 shows a qualitative comparison of one of the models 
($\sigma =0.1$) with the structuring of the ``jets'' in He 2-90 (figure 3 in
Sahai \& Nyman 2000). As stated above, the peak densities depend on model
assumptions and, despite the fact that we have not attempted to specifically
reproduce this nebula, the similarity is outstanding. Thus, although the
images actually show features that resemble supersonic, fast jets, their 
observed properties, and our model, indicate that these features could have 
a different nature. If this is the case, the velocity of the knots should 
be close to that of the AGB wind, and the low radial velocity observed by
Guerrero \etal (2001) is compatible with our model. 

Since jets are actually formed by magnetic collimation, the jet 
interpretation of Sahai \& Nyman (2000) and Guerero \etal (2001) has a 
logical link with our results, but there is one important difference. This 
key difference is that the gas in our case is $only$ collimated but $not$ 
accelerated. The collimation of these outflows is solely due to the hoop 
stress of the toroidal field, but their acceleration require an additional 
process. For instance, the acceleration of a jet-like flow in some 
planetaries is achieved after the magnetized wind passes throught a reverse
shock and then is subsequently pinched at the polar axis (R\'o\.zyczka \& 
Franco 1996). The case of He 2-90 is probably different since, although 
magnetic collimation is most likely present, the acceleration mechanism to 
create a ``supersonic fast jet'' seems to be absent. Note that collimation
without acceleration can occur in a freely outflowing magnetized wind.

It is also relevant to ask if the shells formed in our models are stable 
against the 2D undular or interchange modes, or the 3D mixed mode, of the
Parker instability (the 3D mixed mode is a combination of the interchange 
and undular modes; see Hughes \& Cattaneo 1987, and Matsumoto \etal 1993). 
The growth rates of these instabilities depend on the field strenght, 
the acceleration of the magnetized plasma, and size
of the magnetically supported regions, but the fastest growing mode 
corresponds to the interchange instability (in this case, the flux tubes
are displaced in the radial direction, following the pressure gradient).
The undular instability, on the other hand, grows at a slower pace and 
distorts the $B$-field lines, creating a wavy structure along the field
lines. For the model configurations considered here, the simulations are
blind to the undular instability (because this branch grows in a plane
perpendicular to the computational domain; see Kim \etal 2000) but they 
should be able to resolve some modes (\ie those with wavelengths larger 
than the grid size) of the interchange instability. The accelerations of the 
flow, however, are very 
small and the resulting growth time scales are much too large. Thus, for
any practical purpouses, the models are effectively stable to these 
perturbances.

In addition, given that the higher pressure intershell regions become ligther
than the shells, the growth of plain Rayleigh-Taylor instabilities in the 
radial direction could become relevant at some stage. Our simulations cannot 
resolve the fine scales of this instability, but a simple analysis indicates 
that the growth times are also large. One cannot define a sharp interface 
between the (magnetic) intershell and (non-magnetic) shell regions (their 
initial densities are equal), but the maximum pressure difference is simply 
equal to $B_{\rm max}^2/8\pi $. Thus, for a shell thickness $L$, an upper 
limit to the pressure gradient in the radial direction is simply 
$B_{\rm max}^2/8\pi L$, and the corresponding upper limit to the acceleration
during shell formation is $a \sim v_a^2/2L$ (where $v_a$ is the average 
Alfv\'en speed inside the shell). The resulting growth time for R-T 
instabilities is then (Shore 1992),
\beq
\tau \sim \left( \frac{\rho_2 + \rho_1}{\rho_2 - \rho_1} \right)^{1/2}
   \left( \frac{L}{a} \right)^{1/2} \geq \left( \frac{\rho_2 + \rho_1}{\rho_2 - \rho_1}
   \right)^{1/2} \left( \frac{2^{1/2}L}{v_a}\right) , 
\eeq
where $\rho_2$ is the shell density, and $\rho_1$ is the intershell density. 
Clearly, this time scale is always large: at the early stages because the 
density contrast is very small, and because the acceleration drops as 
the density difference grows at later stages. Thus, the growth time of RT 
instabilities is 
always larger than the time scale for shell formation.

In summary, the dynamical effects of a solar-like magnetic cycle 
operating in late AGB stars have been explored with MHD models. Although 
a number of simplifying assumptions have been made, these 2(1/2)D numerical
simulations are able to succesfuly reproduce for the first time the main
observed properties of the HST rings. One obvious limitation of our models
is that we are unable to include the dynamical effects of dust grains, and
cannot explore the magnetic counterpart of the Simis \etal (2001) models. 
Further studies with less restrictive assumptions may shed more light on the
range of MHD effects that may operate in these type of objects.

{\bf Acknowledgments}
It is a pleasure to thank Yervant Terzian for his helpful discussions 
and encouragement about this topic during his visit to Ensenada. We also
thank Jack Thomas for pointing out the recent work on dynamo amplification in
AGB stars. We warmly thank M. Guerrero, L.F. Miranda, Y-H. Chu, M. Rodriguez 
and R.M. Williams for sharing their results with us prior to publication. 
The comments and criticisms of the referee are also gratefully
acknowledged. As usual, we also thank Michael L. Norman and the Laboratory for
Computational Astrophysics for the use of ZEUS-3D. The computations were
performed at Instituto de Astronom\'{\i}a-UNAM. This work has been partially
supported by grants from DGAPA-UNAM (IN130698, IN117799 \& IN114199) and
CONACyT (32214-E).

\begin{center} List of References \end{center}
\begin{description}

\item Balick, B., Wilson, J. \& Hajian, A. R. 2000,  AJ, 121, 354
\item Begelman, M. \& Li, Z. Y. 1992, ApJ, 397, 187
\item Black, D. C. \& Scott, E. H. 1982, ApJ, 263, 696
\item Blackman, E. G., Frank, A., Markiel, J. A., Thomas, J. H. \& Van 
Horn, H. M. 2001, Nature, 409, 485
\item Clarke, D. A. 1996, ApJ, 457, 291
\item Cox, P. 1997, in IAU Symp. No 180 ``Planetary Nebulae'', eds. H. 
J. Habing \& H. J. G. L. M. Lamers, (Kluwer Academic Publishers, 
Dordrecht), 139
\item Franco, J., Ferrini, F., Ferrara, A. \& Barsella, B. 1991, ApJ,
366, 443
\item Habing, H. J. 1996, A\&A Rev.,  7, 97
\item Hrivnak, B. J., Kwok, S. \& Su, K. Y. L. 2001, AJ, 121, 2775
\item Hughes, D. W. \& Cattaneo, F. 1987, Geophys. Astrophys. Fluid
Dyn., 39, 65
\item Garc\'{\i}a-Segura, G., Langer, N., R\'o\.zyczka, M., \&
Franco, J. 1999, ApJ, 517, 767
\item Guerrero, M., Miranda, L.F., Chu, Y-H., Rodr\'{\i}guez, M. \& 
Williams, R.M., 2001, pre-print.
\item Kim, J., Franco, J., Hong, S. S., Santill\'an, A. \& Martos, M.
2000, ApJ, 531, 873
\item Kwok, S., Su, K. Y. L., Hrivnak, B. J.  1998, ApJ, 501, L117
\item Lazarian, A. \& Vishniac, E. 1999, ApJ, 517, 700
\item Mac Low, M.-M., Norman, M. L., Konigl, A., \& Wardle, M. 1995, 
ApJ, 442, 726
\item Matsumoto, R., Tajima, T., Shibata, K. \& Kaising, M. 1993, ApJ,
414, 357
\item R\'o\.zyczka, M. \& Franco, J. 1996, ApJL, 469, L127
\item Sahai, R., \& Nyman, L.-ÅA.  2000, ApJ, 538, L145
\item Shore, S. N. 1992, {\em An Introduction to Astrophysical 
Hydrodynamics}, (Academic Press, London)
\item Smith, E.J., Tsurutani, B.T., \& Rosenberg, R.L. 1978, J. Geophys. 
Res., 83,
717
\item Soker, N. 2000, ApJ, 540, 436
\item Stone, J. M. \& Norman, M. L. 1992, ApJS, 80, 753
\item Terzian, Y., \& Hajian, A. R. 2000, {\em Asymmetrical Planetary 
Nebulae II: From Origins to Microstructures}, ed. J. H. Kastner, N. 
Soker, S. A. Rappaport, A.S.P. Conf. Series, 199, 33
\item  Wilcox, J., \& Ness, N. 1965, J. Geophys. Res., 70, 1233

\end{description}

\newpage 
\begin{center} {\bf Figure Captions}
\end{center}
\begin{description}

\item Figure 1. 
Equatorial cut for the run with $\sigma = 0.01$ showing the effects
of a cyclic polarity inversion of the AGB stellar magnetic field. 
Each density peak 
correspond to 1,000 yr,  half of the full period.
Units are in c.g.s., except for the radial velocity (\kms).
 
\item Figure 2.  Emission measure of four different runs
projected with a tilt of  $45^{\circ}$ from the plane of the sky.
The panels correspond to:
(top-left) $\sigma = 0.001$, (bottom-left) $\sigma = 0.01$, 
(top-right) $\sigma = 0.05 $, (bottom-right) $\sigma = 0.1$.
Each concentric shell correspond to 1,000 yr. Note the piled-up
blobs at the polar axis for the most magnetized runs, resulting from the
self-collimation of the AGB wind.

\item Figure 3. 
Comparison between the polar region of the model with
 $\sigma=0.1$ (projected
without tilt on the plane of the sky) (top), with
figure 3 by Sahai \& Nyman (2000) of  He 2-90 (bottom). In their figure,
the relative intensity of the jet is plotted as a function of radius  
showing the radial offsets of the knots from the center 
(thick curve: southeast jet, thin curve: northwest jet). 
Both plots are logarithmic in the Y axis, and linear in the X axis.

\end{description}

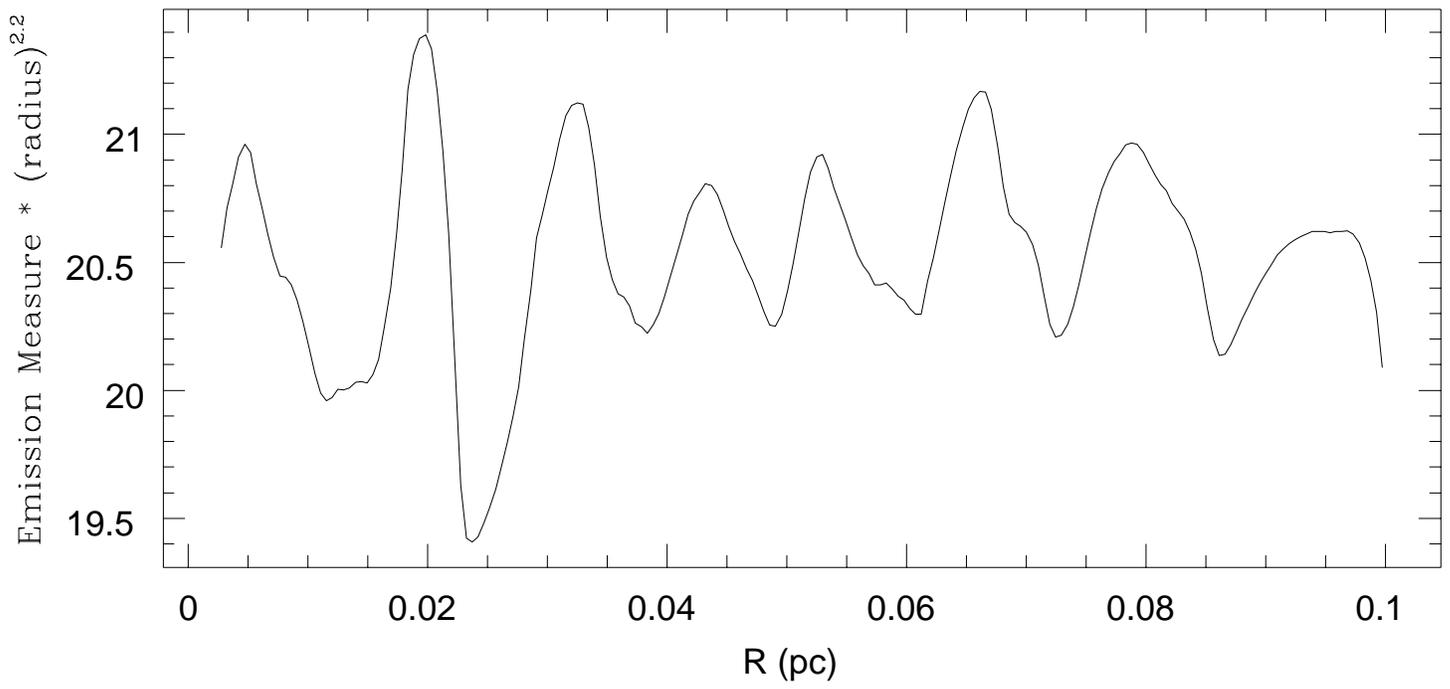
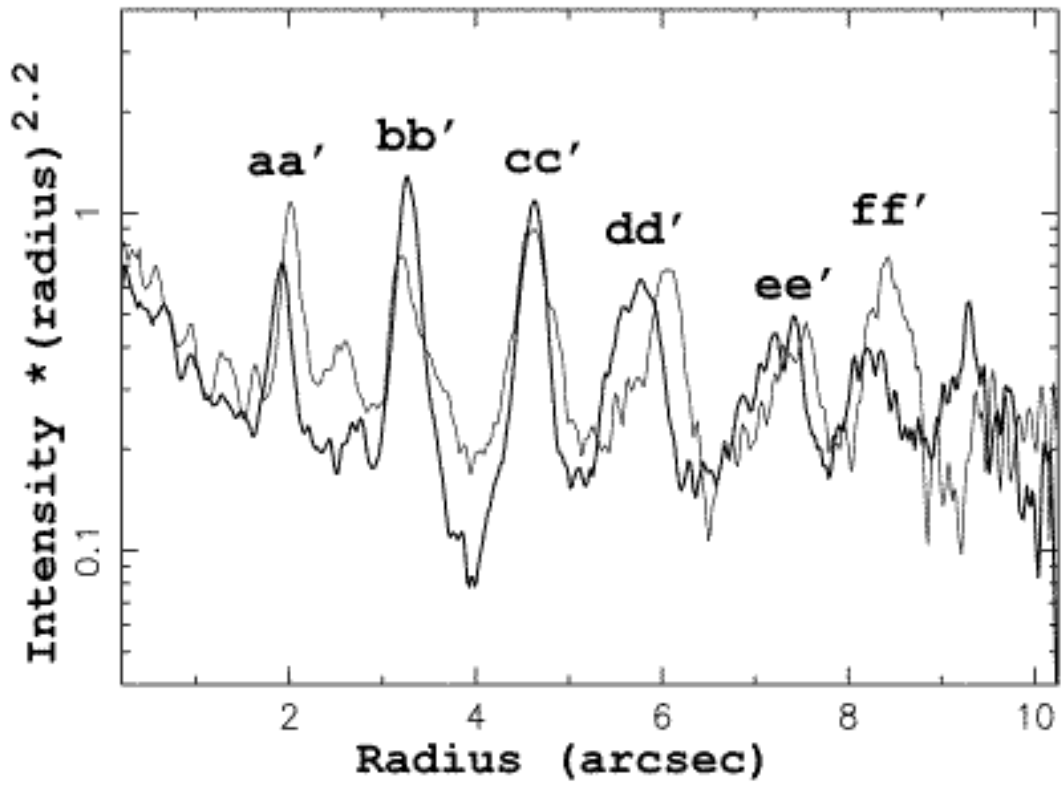
\end{document}